\documentclass[12pt,preprint]{aastex}
\usepackage{amsmath}
\usepackage{natbib}
\usepackage{graphicx}
\usepackage{rotating}

\def\kmps{{km\,s$^{-1}$}}

\newcommand{\kms}{\>{\rm km}\,{\rm s}^{-1}}
\newcommand{\masyr}{\>{\rm mas}\,{\rm yr}^{-1}}
\newcommand{\uasyr}{\>\mu{\rm as}\,{\rm yr}^{-1}}
\newcommand{\muw}{\mu_{\rm W}}
\newcommand{\mun}{\mu_{\rm N}}

\shortauthors{Yang et al.}

\begin{document}

\title{Could the Magellanic Clouds be tidal dwarves expelled from a past-merger event occurring in Andromeda?}

\author {
       Y. Yang\altaffilmark{1,2}                     
   \and F. Hammer\altaffilmark{2}                  
}
\altaffiltext{1}{Key Laboratory of Optical Astronomy, National Astronomical Observatories, Chinese Academy of Sciences, 20A Datun
Road, Chaoyang District, Beijing 100012, China}
\altaffiltext{2}{GEPI, Observatoire de Paris, CNRS, Universit\'e Paris Diderot; 5 Place Jules Janssen, Meudon, France}

\begin {abstract}
The Magellanic Clouds are often considered as
outliers in the satellite system of the Milky Way because they are irregular and gas-rich galaxies.  
From their large relative motion, they are likely from their first pass near the Milky Way, possibly originating from another region of the Local Group or its outskirts. M31 could have been in a merger stage in its past and we
investigate whether or not the Large Magellanic Cloud could have been a tidal dwarf
expelled during this event. Such an
hypothesis is tested in the frame of present-day measurements and uncertainties of the
relative motions of LMC and M31. Our method is to trace back the LMC trajectory using several thousands of different configurations that sample the corresponding parameter space. 

We find several configurations that let LMC at 50 kpc from M31, 4.3 to 8 Gyrs ago , depending on the adopted shape of the Milky Way halo. For all configurations, the LMC velocity at such a location is invariably slightly larger than the escape velocity at such a radius. The preferred solutions correspond to a spherical to prolate Milky Way halo, predicting a transversal motion of M31 of less than 107  \kmps and down to values that are close to zero. We conclude that from present-day measurements, Magellanic Clouds could well be tidal dwarves expelled from a former merger events occurring in M31.
\end {abstract}


\section {INTRODUCTION}

The origin of the Magellanic Clouds, 
as well as the nearby satellite galaxies of the Milky Way (MW),
is still a matter of debate  
\citep{2007ApJ...668..949B,2009arXiv0907.5207P,2009ApJ...700..924K,
2008ApJ...680..287M,2009ApJ...697..269M}. 
The discussion is motivated by the accurate determination of the Clouds' proper motions
that were carried out from the Hubble Space Telescope (HST) observations
by \citet{2006ApJ...638..772K,2006ApJ...652.1213K}. 
The total velocity of the Large Magellanic Cloud (LMC) 
in the Galactocentic coordinate is claimed be $378$ \kmps\
(a transverse velocity of $v_{\rm tan}=367$~\kmps\ and a radial velocity of $v_{\rm rad}=89$~\kmps).
Although the revised analysis by \citet{2008AJ....135.1024P} 
decreases the transverse velocity to $346$~\kmps, both of the results 
from HST data are significantly higher than the previously adopted value, 
i.e., 281~\kmps\ \citep{2002AJ....124.2639V}. At such a high speed, 
LMC may approach the escape velocity at its distance to the MW and the orbital angular momenta 
is comparable and nearly perpendicular 
to the angular momentum of the MW  disk \citep{2009ApJ...700..924K}. 
This may argue for a first passage of  the Magellanic Clouds near the Milky Way 
\citep[see e.g.,][]{2007ApJ...668..949B}. 
The fact that their morphologies and gas content is at odd with other satellites also 
suggest that they recently falls to the Milky Way from the outskirts of the Local Group 
\citep{2006AJ....132.1571V}.

\citet{2009ApJ...700..924K} construct a model for the Local Group,
including Andromeda (M31), the MW and the LMC. 
By solving the equations of motion, they find that M31 may have affected
the orbit of LMC at a distance of 500\,-\,700 kpc about 5 Gyrs ago.
Although \citet{2007ApJ...668..949B} did not investigate the origin
of LMC by their models, one interesting orbit of LMC is worth to mention here.
In their Fig.~14, the orbit under the model of prolate MW halo turns close to
the direction of M31.  
The proposition that Magellanic Clouds may originate from M31 was firstly made by 
\citet{1989MNRAS.240..195R},
\citet{1994AJ....107.2055B} and \citet{1992ApJ...386..101S}.

This {\it Letter} revisits this proposition in the frame of the recent discoveries of large scale structures surrounding M31 suggesting a very tumultuous past history for this galaxy \citep{2001Natur.412...49I,2004MNRAS.351..117I,2008ApJ...685L.121B}. We are still lacking of a complete model of M31 outskirts although many of its properties are consistent with a past major merger (\citealt{2007ApJ...662..322H,2010MNRAS.401L..58B}). If true, such a merger should have been gaseous rich enough to allow the reformation of the significant M31 disk (\citealt{2005A&A...430..115H,2009A&A...507.1313H}). During such events gas-rich dwarf galaxies may be formed from material liberated by the collision \citep[][and references therein]{2000ApJ...543..149O}. 
It is natural to wonder whether or not some tidal dwarf galaxies may have been ejected close to the orbital plane of the hypothetic merger, which is indeed defined by the actual M31 disk. A significant part of the ejected material have angular momentum within small angles from the orbital angular momentum. The M31 disk is seen almost edge-on from the Milky Way, suggesting that the Milky Way is located close to the orbital plane of the debris ejected from a major merger of M31.  This may lead to a fully new interpretation of the Magellanic Clouds, that could be tidal dwarves, as massive and concentrated debris lying in a tidal tail ejected during a past event in M31, in the direction of the Milky Way. In fact, \citet{Hammeretal.2010} propose a major model for the formation of M31 which reproduces most of its properties including those of its haunted halo; for {\it some} solutions, a significant amount of matter is predicted to be ejected from the merger in the direction of the Milky Way.

The goal of this {\it Letter} is to test whether the Magellanic Clouds could have been tidal dwarves 
ejected during a past major merger occurring at the M31 location.  
The robust measurements of LMC proper motion give a very strong constrain 
on its origin by inverting its past trajectory \citep[e.g.,][]{2009ApJ...700..924K}. 
On the other hand, there is a large uncertainty in the determination of 
the tangential motion of M31, up to $\pm 150 \kms$, 
\citep[see][]{2001ApJ...554..104P,2005ApJ...633..894L, 
2008ApJ...678..187V}. Taking into account all uncertainties, we 
 investigate the possible trajectories of the LMC and
whether or not it could have approached M31 down to 50 kpc. 
We solve the equations of motion in a dynamical model including MW, M31 and LMC, 
and throughout the paper, all the 3D coordinates, velocities are quoted
in the Galactocentric frame that is centered on the MW \citep{2002AJ....124.2639V}.
We adopt the concordance cosmological parameters of
$H_0\!=\!70$ km s$^{-1}$ Mpc$^{-1}$, $\Omega_M\!=\!0.27$ and
$\Omega_\Lambda\!=\!0.73$.

\section{Analysis}
\begin{figure}[!t]
\epsscale{0.5}\plotone{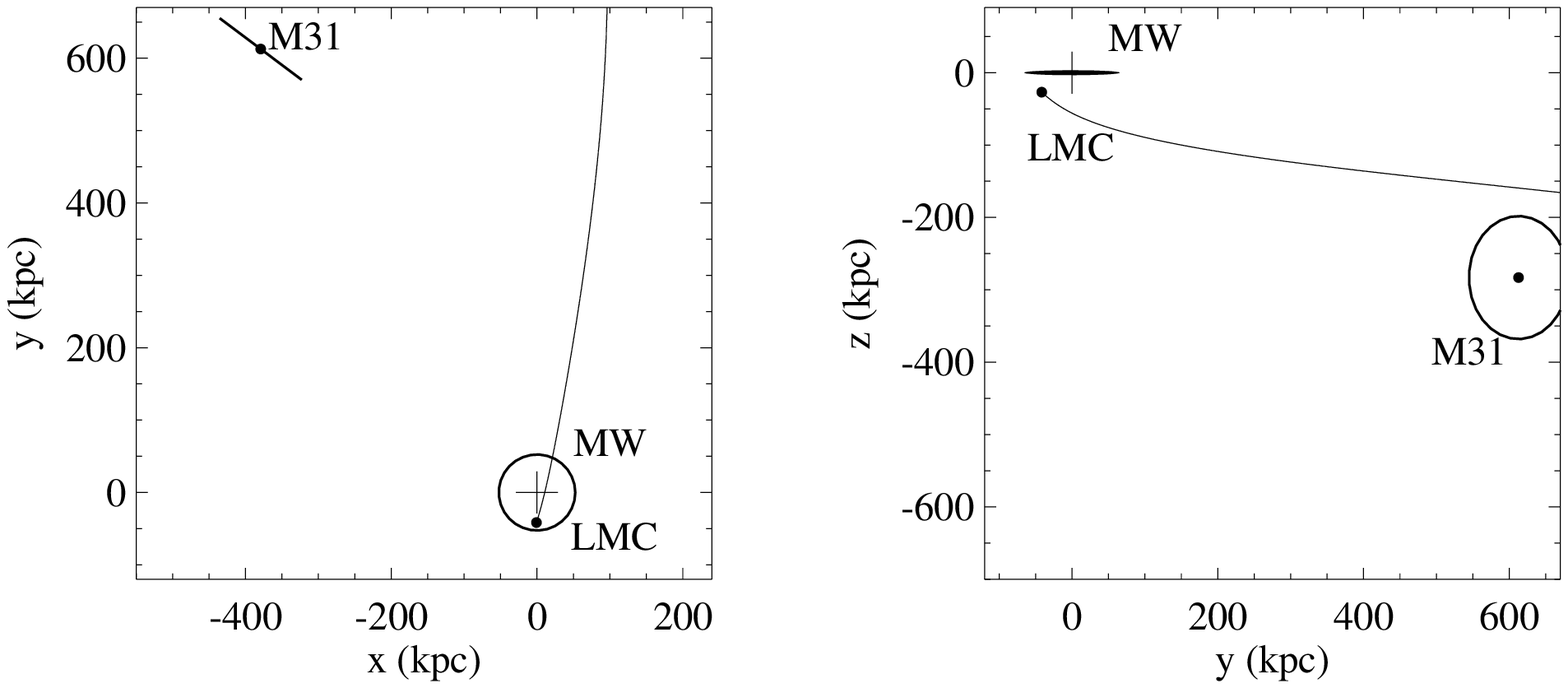}\\
\vspace{0.5cm} 
\epsscale{0.35}\plotone{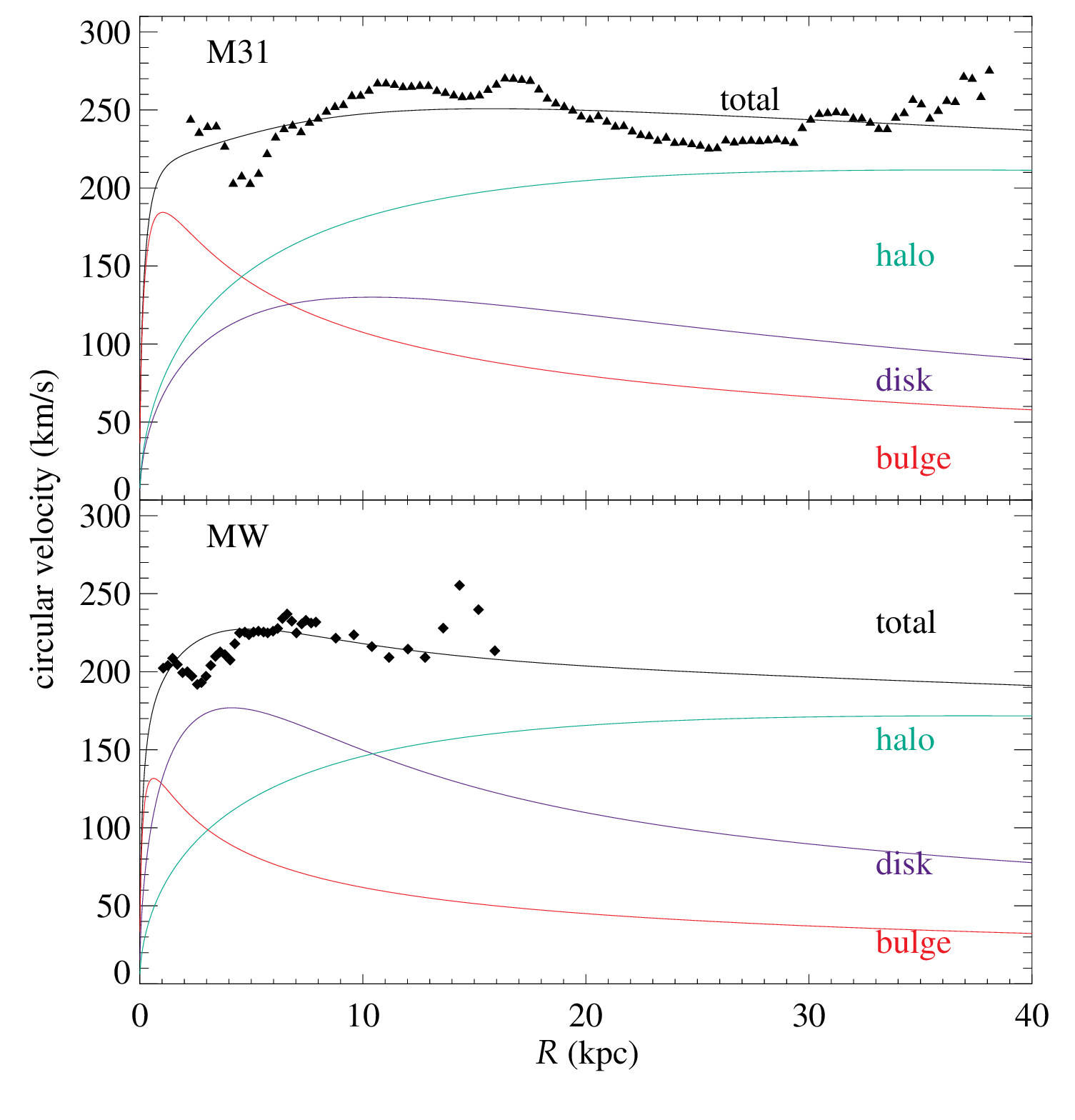}
\caption{{\it Upper panels}: 3D positions of the MW, M31, LMC.
The left panel shows the projection in x-y plane 
while the right panel in y-z plane.
The solid lines indicate
a possible unbound trajectory of LMC in the past.
{\it Middle panel}: the rotation curves of M31. 
The triangles are the measured rotation curve from HI observation
\citep{2009ApJ...705.1395C}. The red, blue, green and black are 
the rotation curve of bulge, disk, halo the total of them, respectively 
(Sect.~\ref{secmethod}). 
{\it Lower panel}:  the rotation curves of MW. The diamonds are from HI observation 
by \citet{1985AJ.....90..454K}.
\label{figpositions}}
\end{figure}

Fig.~\ref{figpositions} shows the 3D positions of the MW, M31 and LMC.
A possible unbound trajectory of LMC in the past is also shown 
by assuming a zero transverse velocity of M31 for our  
model described below (Sect.~\ref{secmethod}). 
This solution is similar to the results presented by \citet{2009ApJ...700..924K}.
Given the fact of large uncertainties in the determination of 
tangential motion of M31, up to $\pm 150 \kms$ (see above text), 
one can expect that M31 could have a $v_x<0$ at present time, 
meaning that M31 was closer to the past trajectory of LMC. 
The 3D velocity of M31 can be linked to its proper motion on sky by 
following the work by \citet{2002AJ....124.2639V}.
We adopt the standard IAU values, i.e., $R_0=8.5$ kpc  and $V_0=220$ $\kms$ 
for the circular velocity \citep{1986MNRAS.221.1023K},
and the solar motion with respect to the local standard of rest is corrected by taking 
$(U_\odot,V_\odot,W_\odot)=(10.0\pm 0.4,5.2\pm 0.6, 7.2\pm 0.4)\, \kms$ 
\citep{1998MNRAS.298..387D}.
The basic data adopted for M31 
are listed in Table~\ref{tbobspar}, 
as well as the data for LMC.  
In the following, we build a dynamical model of MW, M31 and LMC, then investigate 
the possible proper motions of M31 and its impact on the LMC origin.

\begin{table}[!t]
\caption{Basic parameters}
\label{tbobspar}
{\small
\begin{tabular}{lcc} \hline \hline
Parameter      &  M31        &  LMC \\ \hline
 $(l,b)$       &  $(121.174,-21.573)$$^{\rm a}$ & $(280.531,-32.523)$$^{\rm b}$ \\
$(\muw,\mun)$ ($\masyr$)  &  {\small to be investigated} &  $(-2.03\pm0.08,0.44\pm0.05)$$^{\rm c}$ \\
$v_{\rm sys}$ ($\kms$) & $-301$$^{\rm d}$ & $ 262.2$$^{\rm b}$ \\
 $D_0$        (kpc) & $770$$^{\rm e}$ &  $50.1$$^{\rm f}$ \\
\hline 
\end{tabular} \\
$^{\rm a}$~NED.
$^{\rm b}$~\citet{2002AJ....124.2639V}.
$^{\rm c}$~\citet{2006ApJ...638..772K}.
$^{\rm d}$~\citet{1999AJ....118..337C}.
$^{\rm e}$~\citet{2008ApJ...678..187V}.
$^{\rm f}$~\citet{2009ApJ...700..924K}.
}
\end{table}

\begin{table}[!t]
\caption{Model parameters}
\label{tbmodel}
{\small
\begin{tabular}{lccc} \hline \hline
 Parameter    & MW & M31 & LMC\\ \hline
$M_{\rm virial}$ ($10^{12}M_{\odot}$)  & 1.0$^{\rm a}$ & 1.6$^{\rm a}$ & --- \\
$R_{\rm virial}$ (kpc) & 258$^{\rm b}$ & 300$^{\rm b}$ & --- \\ 
$c$$^{\rm c}$  & 15 & 18 & --- \\ 
$M_{\rm baryon}$ ($10^{10}M_{\odot}$) & 5.6$^{\rm d}$ & 10.9$^{\rm d}$ & 2.0 \\
B/T$^{\rm e}$  & 0.15 & 0.3  & --- \\
$a_{\rm b}$ (kpc)  & 0.62 & 1.0 & --- \\
$r_{\rm d}$ (kpc) & 2.3$^{\rm d}$ & 5.8$^{\rm d}$ & ---\\  \hline
\end{tabular}  \\ 
$^{\rm a}$~Data are from \citet{2007ApJ...668..949B}.
$^{\rm b}$~Data are from \citet{2002ApJ...573..597K}.
$^{\rm c}$~The concentration of the NFW profile.
$^{\rm d}$~Data are from \citet{2007ApJ...662..322H}.
$^{\rm e}$~B/T is defined as the mass ratio of bulge to the total baryon mass. 
}
\end{table}

\subsection{Dynamical model}
\label{secmethod}
Following \citet{2007ApJ...668..949B} and \citet{2009ApJ...700..924K}, we constructed a model of  
the MW, M31 and of the LMC, the latter being considered as a point mass with a total mass 
of $2\times 10^{10}M_{\odot}$.
For both the MW and M31 we adopt a model consisting of a NFW halo 
\citep{1997ApJ...490..493N},  
a Hernquist bulge \citep{1990ApJ...356..359H} and an exponential disk.
Then the total gravitational potential of the galaxy model is the sum-up 
of the three components \citep{2007MNRAS.377...50H,2009MNRAS.392L..21S}:
\begin{eqnarray}
\label{eqphis}
\phi(r)  & =  & \phi_{\rm b}(r) + \phi_{\rm d}(r) + \phi_{\rm h}(r), \\
\phi_{\rm b}(r) & = & -\frac{GM_{\rm b}}{r+a_{\rm b}}, \\
\phi_{\rm d}(r) & = &-\frac{GM_{\rm d}(1-e^{-\frac{r}{R_d}})}{r}, \\
\phi_{\rm h}(r) & = & -\frac{GM_{\rm vir}/r_{\rm s}}{\ln(1+c)-c/(1+c)}\frac{\ln(1+r/r_{\rm s})}{r/r_{\rm s}}, 
\end{eqnarray}
where the subscript b, d, and h denote bulge, disk and halo, respectively;
$a_{\rm b}$ the scale length of Hernquist profile;
$R_d$ the scale length of disk; 
$c$ and $r_{\rm s}$ the concentration and scale length of the NFW profile.
The parameters of the model are summarized in Table~\ref{tbmodel}. 
The models are required to match the rotation curves of the MW and M31, 
respectively, see Fig.~\ref{figpositions}.

The equation of motion for each object can be written as:
\begin{equation}
\frac{d^2}{dt^2} \vec{r}_i = 
\frac{\partial}{\partial \vec{r}_i} \sum_{j\ne i}
\phi_j[\mid\vec{r}_i-\vec{r}_j\mid].
\label{eqmotion}
\end{equation}
Then the trajectories of each object can be solved numerically 
using the standard method for N-body simulation in barycentric frame. 
By choosing a small time step for the integration, it provides us an
accuracy down to 0.1\% over 10 Gyr, which is precise enough for our discussions.

As mentioned in the introduction, nonspherical halo of MW may have impacted
to the trajectory of LMC as seen in \citep{2007ApJ...668..949B}.  The cases of non-spherical MW halo
can be studied by replacing ${r}$ in $\phi_{\rm h}(r)$ by 
$r = \sqrt{R + {z^2}/{q^2}}$, where the cylindrical polar coordinates
is adopted, $q$ characterizes the axis ratio of halo potential
\citep{2007MNRAS.377...50H,2007ApJ...668..949B}. For $q>1$ we refer to a prolate halo
while $q<1$ to an oblate halo.

\subsection{Results}
We uniformly sampled $\sim$2000 possible M31 proper motions with the amplitude of [0,0.12]\,$\masyr$
and the orientation [0,360] degrees on the sky. 
We define a reasonable solution by 
searching when the minimal distance between LMC and M31 can be less than 50 kpc, enough close to be consistent with material ejected from an ancient merger, during the last 10 Gyrs.
We do find a group of solution by using the LMC proper motion from \citet{{2006ApJ...638..772K}}. 
The solution of M31 proper motion for a spherical MW halo is:
\begin{eqnarray}
\label{eqms}
\mu_{\rm W}&=&-62\pm 18\uasyr,  \\
\mu_{\rm N}&=&\nonumber-25\pm 13\uasyr,
\end{eqnarray}
where the error bar accounts for the error of LMC proper motions, and $\mu_{\rm W}$, $\mu_{\rm N}$ are quoted in the equatorial system as usually used.
It corresponds to $v_{\rm rad}$=$-128\kms$ and $v_{\rm tan}$=102$\kms$ for M31 relative to the MW.
The averaged time since LMC was ejected from M31 is $5.5\pm1.4$~Gyrs ago. 
Table 3 summarises the results after assuming different values for the axis ratio of the Milky Way potential. In this Table $v_{\rm tan}$, $T_{\rm travel}$ and $v_{\rm 50}$ are averaged values for all the trajectories that put the LMC at 50 kpc from M31 at lookback times indicated by $T_{\rm travel}$. At such a distance from M31 and for all solutions, the relative velocity of LMC to M31 was slightly higher than the escape velocity which is 408 $\kms$, consistently with expectations for material ejected from M31.  Note that we define the escape velocity when an object arrives to the intergalactic space, i.e., 600 kpc from M31, between the Milky Way and M31.

\begin{figure}[!t]
\centering
\begin{tabular}{cc}
\includegraphics[width=9cm]{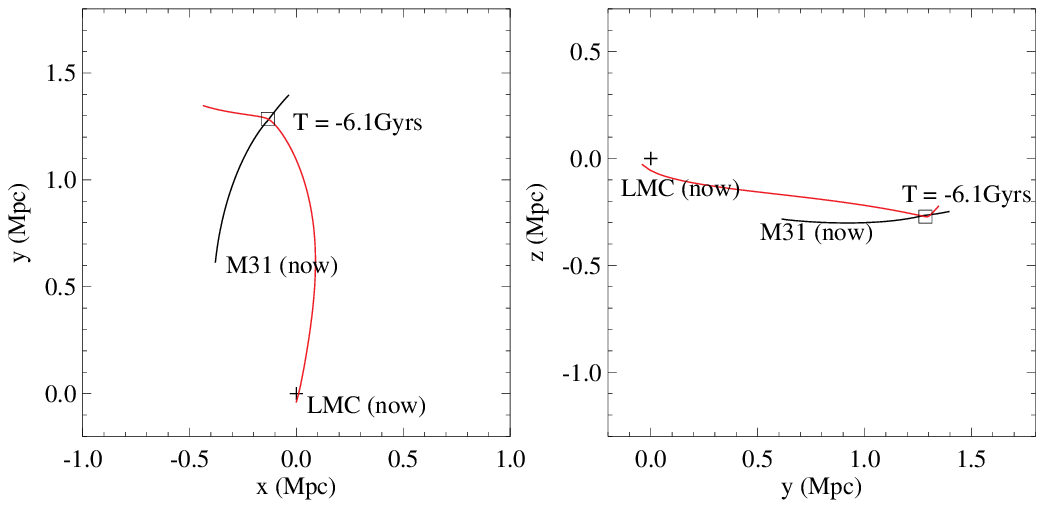}~~~\includegraphics[width=4.3cm]{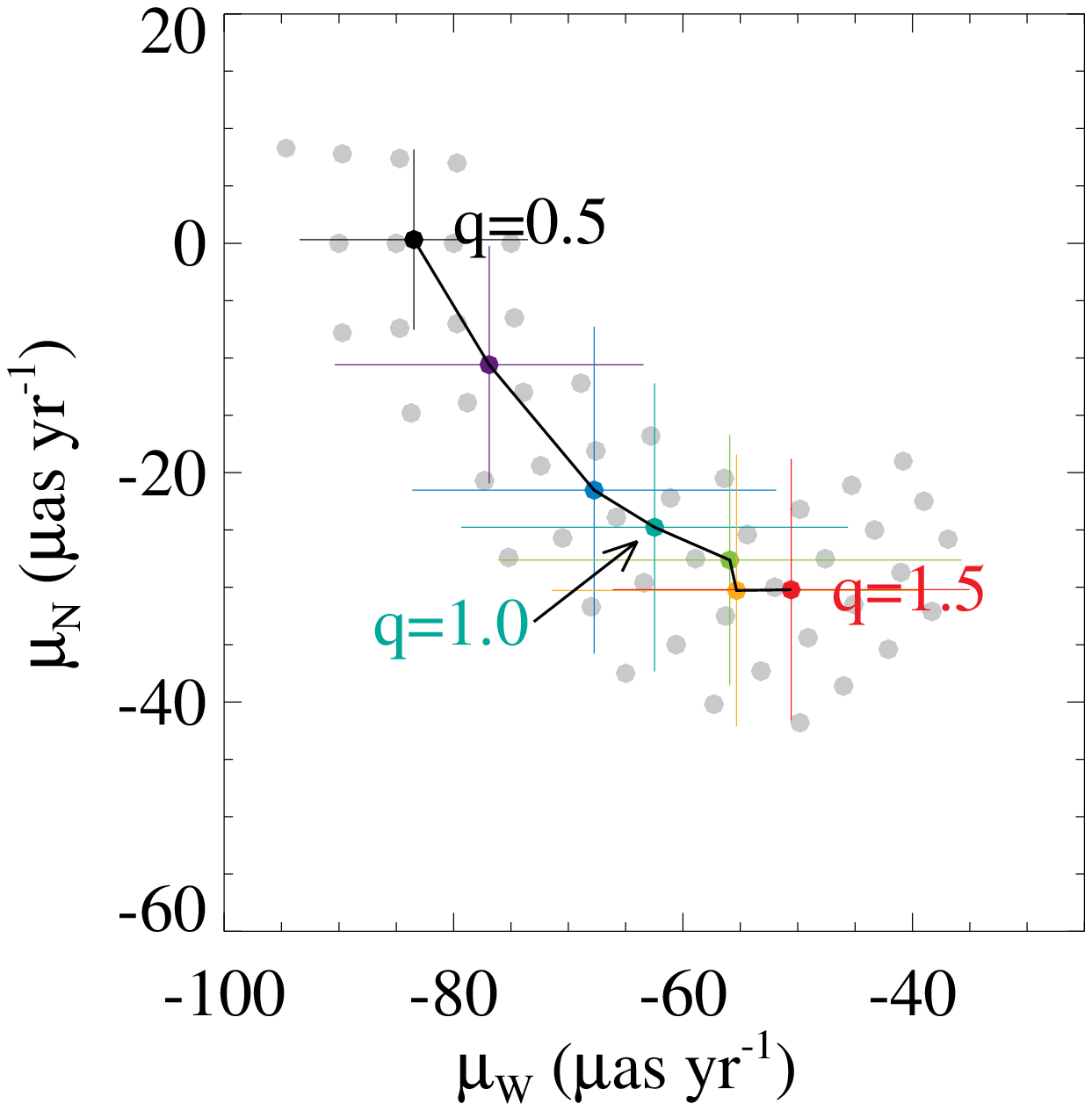}
\end{tabular}
\caption{{\it Left and Middle panels}: 
Trajectories of LMC (red) and M31(black) for the mean solution of a spherical MW halo (Eq.~\ref{eqms}).
The open boxes indicate the time when LMC is closest to M31.
{\it Right panel}: Possible M31 proper motions that satisfy our hypothesis.
The gray dots are all the possible solutions, including by varying of MW shape, i.e., $q$, 
and accounting for the measurement errors of the LMC proper motion.
The color dots connected by a line indicate the mean value for different $q$. 
The bars attached to each point indicate the solution region for each $q$. 
}
\label{figtrajes}
\end{figure}

In Fig.~\ref{figtrajes} we show the trajectories of M31 and LMC for the mean solution ($q=1$).
In the right panel of Fig.~\ref{figtrajes} we show the solutions of M31 proper motion varying with $q$ that correspond to the average of all successful solutions. We have
scanned a region of $q=[0.5,1.5]$ with span of 0.2, i.e., from oblate to prolate (see also Table~\ref{tbsolus}).  Note that $q$ is used in potential space, therefore the halo shape in density space would be even more extreme for both prolate and oblate ones.

\begin{table}[t]
\caption{Possible Solutions}
\label{tbsolus}
{\small
\begin{tabular}{c|cccccc} \hline \hline
$q$$^{\rm a}$ & $\mu_{\rm W}$ & $\mu_{\rm N}$ & $v_{\rm rad}$$^{\rm b}$ & $v_{\rm tan}$$^{\rm b,c}$ & $T_{\rm travel}$ & $v_{\rm 50}$$^{\rm d}$  \\
       &  \multicolumn{2}{c}{($\uasyr$)} & $\!\!\!\!\!\!$($\kms$) & ($\kms$) & (Gyr)&  $\!\!\!\!$($\kms$)\\ \hline
0.5 &$-83\pm 10$ & $~~\,01\pm 08$ &  $-127$ & $   194\pm 44$ &$ -4.3\pm 0.5 $ & 432  \\
0.7 &$-77\pm 14$ & $  -10\pm 10$ &  $-128$ & $   160\pm 56$ &$ -4.5\pm 0.6 $ & 428  \\
0.9 &$-68\pm 16$ & $  -21\pm 14$ &  $-128$ & $   124\pm 53$ &$ -5.1\pm 1.1 $ & 423  \\
1.0 &$-62\pm 18$ & $  -25\pm 13$ &  $-128$ & $   106\pm 63$ &$ -5.5\pm 1.4 $ & 421  \\
1.1 &$-56\pm 20$ & $  -28\pm 11$ &  $-129$ & $    89\pm 68$ &$ -6.2\pm 1.9 $ & 418  \\
1.3 &$-56\pm 16$ & $  -30\pm 09$ &  $-129$ & $    89\pm 48$ &$ -7.8\pm 2.2 $ & 417  \\
1.5 &$-52\pm 15$ & $  -32\pm 09$ &  $-129$ & $    80\pm 46$ &$ -7.2\pm 2.3 $ & 417  \\
\hline
\end{tabular}
}\\
$^{\rm a}$~The shape parameter of MW halo.
$^{\rm b}$~The velocities are given relative to the Milky Way.
$^{\rm c}$~The error bars actually delineate the solutions regions.
$^{\rm d}$~The velocity of LMC at 50 kpc to the M31 center.
\end{table} 
\vspace{1cm}

\section{Discussion and Conclusion}

Could the Magellanic Clouds be ejected tidal dwarves from a previous major merger occurring at the M31 location? In this letter we simply demonstrate that it could be the case, within a reasonable range of parameters for both the Milky Way and M31. In our study we did not consider the effect of the Small Magellanic Cloud (SMC) to LMC as the former has a much smaller mass than the latter, and as such, our study of the LMC motion can realistically apply to the motion of both Magellanic Clouds together. We have also considered the possible impact of the gravitational potential of both the Virgo cluster and the Great Attractor \citep[see e.g.,][]{1988ApJ...326...19L}, which are modelled by giant halos of NFW profile and find that they affect very marginally the LMC trajectories. The model presented here is possibly static assuming that only the M31 mass could have increase during the last 8 Gyr, through a major merger, thus neglecting some possible mass accretion of minor mergers. Assuming a smaller mass of M31 in the past (2 times less massive) would generate a larger travel time for the LMC (8 Gyr instead of 5.5 Gyr for our $q=0$ model).

Besides this, we show that such an hypothesis is possible, although it does not demonstrate that it is indeed the case. To go beyond requires an estimate of several quantities, especially the tangential velocity of M31 and the axis ratio of the Milky Way potential. \citet{2008ApJ...678..187V} have estimated tangential velocity of M31 to be $\mu_{\rm W}= -22\pm 12\,\uasyr$ and  $\mu_{\rm N}= -11\pm 10\,\uasyr$, i.e. in the same direction than assumed  in Eq.~\ref{eqms}), but with smaller amplitude.  These estimates were based on the assumption that M31 satellites  follow the motion of M31 through space. We have tested a null hypothesis for the tangential velocity of M31 and find that it implies a very prolate Milky Way halo with $q= 1.7$. However the \citet{2008ApJ...678..187V}  assumption may not hold in the case of an  major merger in the past history of M31 because the orbital motions of its satellite system could be much more chaotic than expected by  \citet{2008ApJ...678..187V}. On the other hand, assuming a spherical halo for the Milky Way leads to values for the M31 proper motion which are larger than what is typically quoted, often on the basis of the timing argument. Possibly, the timing argument has to be re-formulated in a scheme for which there were more than 2 bodies with a mass similar to the Milky Way, 6 Gyr ago \citep[see e.g.,][]{Hammeretal.2010}.

It is wiser to test our hypothesis by considering observable parameters that are not assuming a specific history for the Local Group satellite system. The result is somewhat troubling. 
First, by tracing back the LMC motion to M31, it is found that its relative velocity to M31 at 50 kpc is slightly above the escape velocity which is quite expected for tidal material ejected from a merger. Second, the travel time to reach the Milky Way is ranging from 4 to 8 Gyrs, depending on the axis ratio of the Milky Way potential (see Table 3). 
\citet{2007ApJ...662..322H} estimated that if M31 have experienced a gaseous rich major merger, it should have occurred 5-8 Gyrs ago on the basis of the age of the M31 disk stars.   Indeed in such an event most stars in the rebuilt disk should have ages slightly smaller than the merger look-back time. 

The origin of the Magellanic Clouds is still an enigma as they are the only blue, gas rich irregular in the immediate outskirts of the Milky Way. Our proposition has the advantage of explaining them in a consistent way, as being originating from the most massive body in the Local Group that show evidences for a very rich merger history. Future measurements of the M31 transverse velocity (possibly with GAIA) may confirm or infirm its validity. Further modeling of both LMC and SMC could be done to verify whether their internal structures (e.g. the LMC bar) and their star formation history \citep[see e.g.,][]{2009AJ....138.1243H} can be reproduced. Important tests of our hypothesis may come from better estimates of the dark matter content of the LMC (could it be a tidal dwarf if it has a total mass as adopted by \citet{2010arXiv1009.0496P}?) and from verifying whether it is consistent with the numerous features found in the outskirts of the Milky Way.


\begin{acknowledgements}
We would like to thank the referee for the helpful comments. This work has been supported by the China-France International Associated Laboratory "Origins" and by National Basic Research Program of China (973 Program), No.~2010CB833000. We warmly thank Jim Peebles for enlightening discussions about the motion and mass of the LMC. We would like to thank Mathieu Puech for helpful discussions. 

.
\end{acknowledgements}

\end {document}